\documentclass[aps,prl,twocolumn,showpacs,amsmath,amsfonts]{revtex4}
\usepackage{graphicx}
\begin{document}

\title{A phenomenological theory of nonphotochemical laser induced nucleation}
\date{\today}

\author{M. Nardone} \affiliation{Department of Physics and Astronomy, University of Toledo, Toledo, OH 43606, USA}
\author{V. G. Karpov} \affiliation{Department of Physics and Astronomy, University of Toledo, Toledo, OH 43606, USA}

\begin{abstract}
Our analysis of the experimental data related to nonphotochemical laser induced nucleation in solutions leads to the inevitable conclusion that the phase transformation is initiated by particles that are metallic in nature.  This conclusion appears paradoxical because the final products are dielectric crystals.  We show that the experimental results are well accounted for by the theory of electric field induced nucleation of metallic particles that are elongated in the direction of the field.  However, new physical and chemical insights are required to understand the structure of the metallic precursor particles and the kinetics of subsequent dielectric crystallization.
\end{abstract}

\maketitle

There is growing experimental support for electric field induced nucleation of solute particles in supersaturated solutions.  First reported by Garetz et. al. \cite{garetz1996}, the phenomenon referred to as nonphotochemical laser induced nucleation (NPLIN) has been observed with both oscillating \cite{garetz1996, garetz2002, sun2008, lee2008, alexander2009, alexander2009a, sun2009} and static \cite{aber2005} fields. The term `nonphotochemical' emphasizes that there is no light absorption; hence, underlying changes in electronic structure capable of chemical reactions  are ruled out.  In all of the above cases, the final product of nucleation was found to be small dielectric particles.

A clear indication that the field is the primary phase change driver is the alignment of nucleated particles along the direction of the applied field (or laser beam polarization).  That phenomenon has led to a type of `polarization switching' wherein the polymorph (crystal structure) of the nucleated crystal can be controlled by applying either linear or circular polarized light \cite{garetz2002, sun2008}.  The underlying mechanism remains an open question with many practical implications \cite{oxtoby2002}.

\begin{figure}[htb]
\includegraphics[width=0.40\textwidth]{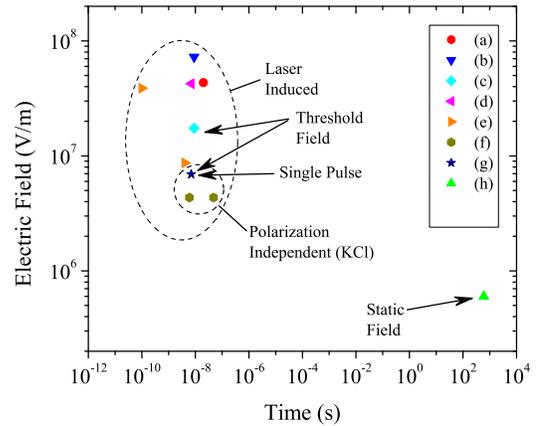}
\caption{Summary of NPLIN data in aqueous solutions of: (a) urea \cite{garetz1996}; (b) KCl \cite {alexander2009}; (c) $\alpha$- and $\gamma$- glycine \cite{garetz2002}; (d) urea \cite{garetz2002}; (e) L-histidine \cite{sun2008}; (f) lysozyme \cite{lee2008}; (g) KCl \cite {alexander2009a}; (h) $\gamma$-glycine \cite{aber2005}.  The ordinate is the peak applied field (unless labeled as the threshold field) at which nucleation occurred within the exposure time on the abscissa.  The typical laser wavelength was $\lambda=1.064$ $\mu$m, except for (e) and (f) at $\lambda=0.532$ $\mu$m. Samples were exposed to numerous laser pulses (except for single pulse exposure in Ref. \onlinecite{alexander2009}) and, typically, less than half of the samples crystallized when irradiated even at the maximum intensity. \label{Fig:data}}
\end{figure}

Our summary of NPLIN data from the literature is presented in Fig. \ref{Fig:data} where the peak laser intensity (or applied field) and exposure time are provided at which crystallization was eventually observed in solutions at various supersaturation levels.  For our purposes, the most important results are that: i) the field reduces the nucleation time by 13 orders of magnitude or even larger; and ii) it can do so at optical frequencies ($\sim 10^{14}$ Hz).  Other observations include the existence of a threshold field, below which nucleation did not occur, and a linear type correlation between the cumulative fraction of samples nucleated and laser intensity \cite{garetz2002, alexander2009}, as well as the solute supersaturation \cite{alexander2009}.

The anomalous strength of the observed field effect can be conveniently expressed in the terms of nucleation barrier $W_0$ that determines the nucleus induction time, $\tau=\tau_0\exp{(W_0/kT)}$.  Here, $\tau_0\gtrsim 10^{-13}$ s is the characteristic atomic vibration or diffusion time, $k$ is Boltzmann's constant, and $T$ is temperature.  The observed reduction of $\tau$ by a factor of $10^{-13}$ requires decreasing the nucleation barrier by approximately $\Delta W(E)=W_0-W(E)\sim 30kT$.  Earlier proposed mechanisms based on the Kerr effect \cite{garetz1996} or isotropic atomic polarizability of dielectric clusters \cite{alexander2009} provided coherent qualitative features but were shown to produce effects on the order of only $10^{-4}kT$, five orders of magnitude below the observations.

Here, we propose a phenomenological theory that shows how NPLIN evolves through nucleation of short-lived, progenitor metallic particles under strong electric field. The metallic particles have gigantic electronic susceptibility compared to dielectric substances and plasma frequency much higher than the laser frequency, thereby allowing the polarization to adiabatically follow the oscillating laser field.  Those properties are unique to metals and correspond, respectively, to the above cited NPLIN results i) and ii).  Consequently, we consider the concept of metallic progenitor particles an experimental fact rather than an hypothesis.  The question of the microscopic structure of the precursor metallic particles remains to be answered and depends on the material system.  A paradoxical feature of our theory is that it introduces metallic particles in processes which are experimentally known to result in the formation of dielectric crystals.

A relevant case of gigantic polarizability is found with a metal needle of length $H$ and radius $R\ll H$. Under an electric field $E$, it will accumulate at its ends opposite charges of absolute value $q\sim EH^2\varepsilon$ corresponding to the dipole moment $p_m\sim qH=EH^3\varepsilon =E\varepsilon \Omega (H/R)^2$, where $\varepsilon$ is the dielectric permittivity of the host material and $\Omega\approx HR^2$ is the particle volume.  For comparison, a dielectric particle of equal volume and characteristic dimension $\Omega ^{1/3}$ will develop, under the same field,  the dipole moment $p_d\sim E\Delta \varepsilon\Omega\ll p_m$, where $\Delta\varepsilon$ is the difference between the dielectric permittivities of the particle ($\varepsilon _p$) and the host ($\varepsilon$).  Hence, the electrostatic energy gain in nucleation, $pE$, is higher for needle-shaped metal particles by a factor of $(\varepsilon /\Delta\varepsilon)(H/R)^2 \gg 1$.

Regarding the dynamic characteristics, we note that the typical metal plasma frequency in the range of $\omega _p\sim 10^{15}-10^{16}$ s$^{-1}$ is much greater than both the laser frequencies used in the NPLIN work and the characteristic dielectric relaxation frequencies corresponding to reorganization and orientation of permanent dipoles, which are all below $\omega _d\sim 10^{11}$ s$^{-1}$.  Therefore, the progenitor metal particles will behave as good metals in the laser field, unlike dielectric particles that would not have time to polarize under the optical frequency laser field.

The above claim of strong energy gain can be made more quantitative in the framework of field induced nucleation (FIN) theory \cite{karpov2008, karpov2010}.  FIN is a recently developed concept of metal phase nucleation in an insulating host under a strong static \cite{karpov2008} or oscillating \cite{karpov2010} field.  We demonstrate that FIN can account for the magnitude of the dramatic field effect observed in the NPLIN experiments.

In general, a particle of volume $\Omega$ in a uniform field $E$ reduces the free energy according to \cite{landau1984},
\begin{equation}\label{eq:elec}
F_E=-\frac{\varepsilon^* E^2}{8\pi}\Omega\quad\mathrm{with}\quad \varepsilon^*=\frac{\varepsilon\Delta\varepsilon}{\varepsilon+n\Delta\varepsilon}.
\end{equation}
Here, the particle polarizability depends on the effective permittivity, $\varepsilon^*$, and the shape of the particle through the depolarizing factor $n$.  A sphere provides $n=1/3$, while for a prolate spheroid of radius $R$ and height $H$, $n=(R/H)^2[\ln(2H/R)-1]$. A metallic particle with $\varepsilon_p\rightarrow\infty$ leads to $n\Delta\varepsilon\gg\varepsilon$, resulting in $\varepsilon^*=\varepsilon/n$, consistent with the intuitive estimate presented above for the dipole moment of a metallic needle.

Insight to the underlying mechanism can be gained by estimating the effective permittivity $\varepsilon^*$ that is required to provide the observed barrier reduction $\Delta W=30kT$.  From Eq. (\ref{eq:elec}), setting $F_E=\Delta W$ yields $\varepsilon^*=8\pi\Delta W/(E^2\Omega)$.  As a rough estimate, a particle of volume $\Omega\sim 1\; \mathrm{nm}^3$ and the experimental $E\sim 10^7$ V/m gives the requirement $\varepsilon^*\sim 10^4$.  In comparison, for a dielectric sphere we have $\varepsilon^*=3\varepsilon(\varepsilon_p-\varepsilon)/(\varepsilon_p+2\varepsilon)$, and using $\varepsilon_p=2.2$ (for KCl) and $\varepsilon=1.8$ (for water at $\lambda=1.064$ $\mu$m) \cite{handbook2007}, yields $\varepsilon^*=0.4$; five orders of magnitude too low.  This crude estimate reveals the unlikelihood that a dielectric particle could provide the necessary barrier suppression.

Nucleation in the presence of an electric field is described by the free energy,
\begin{equation}\label{eq:free}
F=A\sigma-\Omega\mu+F_E,
\end{equation}
where $A$ is the nucleus surface area, $\sigma$ is the coefficient of surface tension, and $\mu$ is the chemical potential difference between the two phases.  For the well known case of spherical nuclei, $A=4\pi R^2$, $\Omega=(4/3) \pi R^3$, and the field-dependent nucleation barrier becomes $W_{sph}=\mathrm{max}\{F(R)\}$,
\begin{equation}\label{eq:sphbar}
W_{sph}=\frac{W_0}{\left(1+E^2/E_0^2\right)^2}\quad\mathrm{with}\quad E_0=2\sqrt{\frac{3 W_0}{\varepsilon^* R_0^3}},
\end{equation}
where $E_0$ is the characteristic field expressed in terms of the classical barrier $W_0=16\pi\sigma^3/(3\mu^2)$ and critical radius $R_0=2\sigma/\mu$. Note that the critical radius in the field is smaller than $R_0$,
\begin{equation}
R_E=\frac{R_0}{\left(1+E^2/E_0^2\right)^2}\label{eq:R_E}\end{equation}

\begin{figure}[htb]
\includegraphics[width=0.35\textwidth]{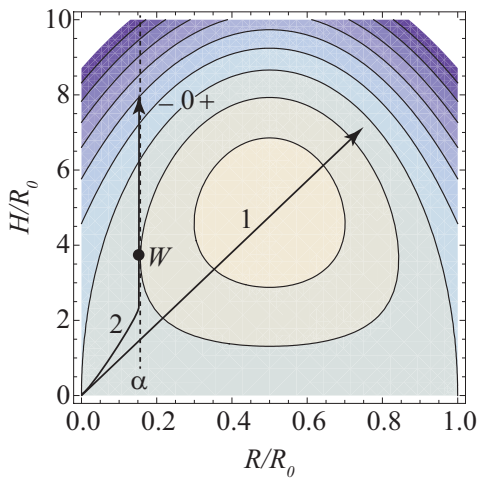}
\caption{Contours of free energy $F/W_0$ from Eq. (\ref{eq:freecyl}); positive and negative regions are separated by the zero contour. The contour spacing is 0.5. Nucleation of elongated particles along path 2 over the barrier $W$ is more efficient than nucleation over the maximum barrier of path 1.  $R/R_0=\alpha$ is the minimum physically reasonable radius. \label{Fig:contour}}
\end{figure}

In FIN, nucleation proceeds through two degrees of freedom by forming needle-shaped particles aligned with the direction of the field (or beam polarization).  They are much more efficient at reducing the electrostatic energy because of their larger dipole moments.  The exact shape of the elongated nucleus is not known, but modeling with either spheroidal or cylindrical particles leads to differences only in numerical coefficients \cite{karpov2008}.  We opt for the mathematically more concise form of a cylindrical nucleus with $A=2\pi RH$ and $\Omega=\pi R^2H$, leading to the free energy of Eq. (\ref{eq:free}) expressed as,
\begin{equation}\label{eq:freecyl}
F_{cyl}=\frac{W_0}{2}\left(\frac{3RH}{R_0^2}-\frac{3R^2H}{R_0^3}-\frac{E^2}{E_0^2}\frac{H^3}{R_0^3}\right).
\end{equation}
Here we have assumed the particle to be metallic with $\varepsilon^*=\varepsilon/n$ and used the approximation $n=(R/H)^2$.  The contour plot in Fig. \ref{Fig:contour} illustrates how the system can lower its free energy more easily by forming elongated particles.

The free energy of Eq. (\ref{eq:freecyl}) seems to suggest that nuclei with $R\rightarrow 0$ are the most favorable.  Realistically, $R$ must be greater than some minimum value determined by extraneous requirements, such as sufficient conductivity to support a large dipole energy or mechanical integrity. Based on data for other types of systems, it was estimated \cite{karpov2008} that a reasonable minimum radius is $R_{min}=\alpha R_0$, where $\alpha\sim 0.1$ is a phenomenological parameter. Lacking more concrete information, we employ the same approximation here.  The free energy in the region $R<R_{min}$ is substantially larger than described by Eq. (\ref{eq:freecyl}), since the energy reducing effect of the electric field cannot be manifested by such thin particles.

As a simplifying approximation we consider nucleation along the path  $R/R_0=\alpha$ (see Fig. \ref{Fig:contour}); alternative paths that start from the origin introduce only insignificant numerical factors.  Then, from Eq. (\ref{eq:freecyl}) the nucleation barrier and critical aspect ratio are,
\begin{equation}\label{eq:cylbar}
W_{cyl}=W_0\frac{\alpha^{3/2}E_0}{E},\quad \frac{H_c}{R_{min}}=\frac{E_0}{\alpha^{1/2}E}\gg 1.
\end{equation}

\begin{figure}[htb]
\includegraphics[width=0.40\textwidth]{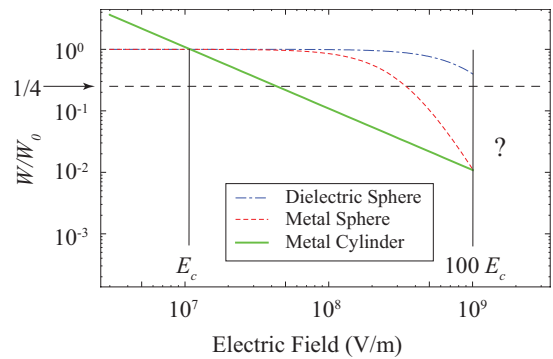}
\caption{Comparison of the field induced barrier suppression from Eqs. (\ref{eq:sphbar}) and (\ref{eq:cylbar}) for a dielectric sphere (dash-dot), metallic sphere (dash), and metallic cylinder (solid).  Experimentally observed barrier reduction $W/W_0=1/4$ (with $W_0=40kT$), shown by the horizontal line, is achieved near $3\times 10^7$ V/m for the metallic cylinder; in the range of experimental data.  The region $E_c<E<E_c/\alpha^2$ (using $\alpha=0.1$) is the effective range of FIN.  Nucleation in the region $E>E_c/\alpha^2$ is uncertain due to the requirement of ultra-small nuclei.  The numerical values are provided in the running text.\label{Fig:barriers}}
\end{figure}

Although the nature of the metallic precursor particles is unknown, with reasonable values of $W_0=1$ eV, $R_0=3$ nm, and $\alpha=0.1$, Eqs. (\ref{eq:sphbar}) and (\ref{eq:cylbar}) imply that the experimentally observed barrier reduction is achieved at a field of $E\sim 10^9$ V/m for dielectric spheres and $E\sim 3\times 10^7$ V/m for metallic cylinders; well within the range of NPLIN data (see Fig. \ref{Fig:data}). In a static field, $\varepsilon\sim 100$ for aqueous solutions and the latter value is reduced to $E\sim 3\times 10^6$ V/m.  The field dependent nucleation barriers for the various scenarios are shown in Fig. \ref{Fig:barriers}.

Comparing Eqs. (\ref{eq:sphbar}) and (\ref{eq:cylbar}) indicates that nucleation of needle-shaped particles is favored when $W_{cyl}<W_0$, resulting in the critical field condition $E>E_c\equiv \alpha^{3/2}E_0$.  The requirement on the aspect ratio $H_c/R_{min}\gg 1$ from Eq. (\ref{eq:cylbar}) implies the upper limit $E<E_0/\sqrt{\alpha}$.  Taken together, FIN is effective in the range,
\begin{equation}\label{eq:range}
1<E/E_c<\alpha^{-2},
\end{equation}
which is clearly indicated in Fig. \ref{Fig:barriers}; $10^7<E<10^9$ V/m for the numerical values mentioned above.  Beyond the upper limit ($E>E_c/\alpha^2$), small nuclei with $R<R_{min}$ are expected.  The nucleation of such small particles can involve other physical aspects that we do not consider here (cf. \cite{wang2008}).  Below the lower range, spherical particles are more probable than cylinders but the field effect is negligible (i.e. minimal barrier suppression).

From the induction time $\tau=\tau_0\exp{(W_{cyl}/kT)}$ and Eq. (\ref{eq:cylbar}), the threshold field is given by,
\begin{equation}\label{eq:threshold}
E_{th}=\frac{W_0}{kT}\frac{E_c}{\ln{\left(\tau/\tau_0\right)}}.
\end{equation}
It provides verifiable predictions in terms of the threshold field dependence on exposure time, temperature and the supersaturation coefficient $\gamma$ through $W_0\propto 1/\ln^2{\gamma}$ \cite{khamskii1969}.  Some care must be taken in interpreting the available data since only the upper limits of the induction times are known and threshold fields were only reported in Refs. \onlinecite{garetz2002} and \onlinecite{alexander2009}.

The above analysis was limited to the nucleation stage of phase transformation. However, post-nucleation growth (or decay) can strongly affect the number of experimentally observed second phase particles.  In general, the post-nucleation processes can be rather complex, including secondary nucleation of the second phase particles on the precursor metallic embryos, structural reconstruction \cite{erdemir2009}, and subsequent particle growth by accretion from the solution. The reconstruction step implies that ``nucleation is, at least, a two barrier process in terms of the thermodynamic potential, in which the first barrier necessary for cluster formation is lower than the main barrier necessary for the transformation of the already formed cluster into a stable crystalline nucleus'' \cite{erdemir2009}; it goes beyond classical nucleation theory and was suggested based on empirical observations.  In what follows, we attempt a qualitative description of how the secondary process depends on the field and solute concentration.

\begin{figure}[htb]
\includegraphics[width=0.37\textwidth]{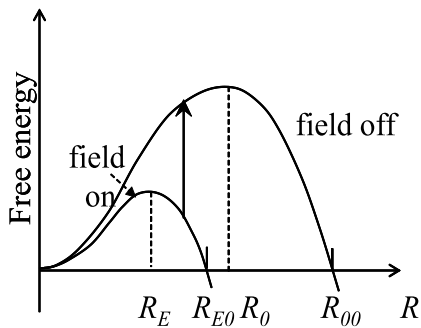}
\caption{Sketch of the particle free energy under zero field and strong electric field. $R_0$ and $R_E$ show the corresponding nucleation barriers, while $R_{00}$ and $R_{E0}$ represent the radii above which the particle becomes energetically favorable. The upward arrow shows the transition that takes place upon field removal, which leads to particle decay. \label{Fig:NuclSol}}
\end{figure}

As illustrated in Fig. \ref{Fig:NuclSol}, a newly nucleated particle remains unstable upon field removal unless its size has grown enough (above $R_{0}$ in Fig. \ref{Fig:NuclSol}) to ensure particle stability (continued growth) in zero field.  Therefore, the field needs to be maintained for a certain time, $\tau_g$, to let a just nucleated particle evolve into the zero-field stability region.  The particle growth rate determines both that time and the number of stable particles found upon field removal.  Assuming the characteristic time of field exposure $\tau _F$, the condition of sufficient growth takes the form $R(\tau _F)>R_0$. 

When a post-nucleation stage of particle formation becomes the bottleneck, then the phase transformation rate will not be exponential in the electric field and material parameters, as would be typical for nucleation processes \cite{kaschiev2003}.  Here, we consider a conceivable scenario inspired by the data in Refs. \onlinecite{garetz2002, alexander2009}.  We assume the growth stage to be the bottleneck that determines the number of particles observed upon field removal, while the characteristic nucleation time $\tau _n\ll \tau _F$ is the shortest of all the processes.  Thus, metal nucleation takes place with certainty during the time $\tau_F$ of field exposure.  Furthermore, we consider the simplest hypothesis that the probability for a particle to grow beyond the stability radius $R_0$ is proportional to the diffusion flux $I$ of molecules from the solute to the particle.  The latter is given by the equation (see e.g. \cite{landau2008}, p. 431)
\begin{equation}\label{eq:flux}
I=4\pi r^2D\,dc/dr=4\pi D(c-c_{0\infty})(R_0-R_E),
\end{equation}
where we have implied a spherical nucleus with radius close to the critical radius $R_0$, $c$ is the solute concentration and $c-c_{0\infty}\equiv\Delta c$ is the solute oversaturation.  Because the practical fields are much lower than $E_0$, it follows from Eq. (\ref{eq:R_E}) that $R_0-R_E\approx 2R_0E^2/E_0^2$, which yields for the number of stable particles,
\begin{equation}\label{eq:Npart}
N\propto E^2\Delta c.
\end{equation}
Note that the dependence in Eq. (\ref{eq:Npart}) can be shown to hold not only for spherical particles [assumed in Eq. (\ref{eq:flux})], but for cylindrical particles as well.  Hence, when the field is sufficient to induce nucleation of metallic precursors ($E>E_c$), the probability of observing crystallization has the dependence of Eq. (\ref{eq:Npart}); as observed in Refs. \onlinecite{garetz2002, alexander2009} and \onlinecite{kozlovskii1966}.

Regarding inquiry into the nature of the metallic progenitor particles, we note that FIN may provide a means by which otherwise chemically unstable substances can persist.  As an example, consider the violently reactive combination of potassium in water with an enthalpy of reaction, $\Delta H\sim 200$ kJ/mol $=2$ eV/atom.  Using $\mu\sim 2$ eV/atom $\sim 4\times 10^9$ J/m$^3$ and $\sigma \sim 20$  eV/atom (for structurally different phases) yields $R_0\sim 20$ $\textrm{\AA}$ and $W_0\sim \mu R_0^3\sim 200$ eV.  Then with $\alpha=0.1$, and $\varepsilon\sim 1$, we obtain $E_c\sim 3\times 10^8$ V/m.  Although this minimum required field is beyond the range of the NPLIN data, the point is that otherwise unexpected substances may become stable in the presence of a sufficiently strong electric field.  In particular, one can assume metal-like particles of some other, more complex composition than pure potassium with considerably lower enthalpy of reaction.

In summary, the most pertinent aspects of the NPLIN data, including the magnitude of the field effect, the polarization dependence, and the threshold field are well described within the framework of FIN.  A qualitative picture of the post-nucleation process leads to a phase transformation rate that depends quadratically on the field and linearly on solute concentration.  The paradoxical requirement is that metallic progenitor particles precede the formation of dielectric crystals.  The structure of the metallic precursors and the kinetics of formation of the dielectric crystals are intriguing and largely open questions.

\end{document}